\documentclass[twocolumn,tighten, twocolappendix]{aastex63}

\newcommand{\argmin}{\mathop{\rm argmin}\limits}

\accepted{April 14, 2020}
\submitjournal{ApJ}

\usepackage{graphicx}
\usepackage{natbib}
\usepackage{longtable}
\usepackage{booktabs}
\usepackage{tabularx}
\usepackage{amsmath,amstext}
\usepackage{amssymb}
\usepackage{apjfonts} 
\usepackage[figure,figure*]{hypcap}

\shorttitle{Super-resolution Imaging of the Protoplanetary Disk HD~142527 Using Sparse Modeling}
\shortauthors{Yamaguchi et al.}

\definecolor{MyDarkBlue}{rgb}{0,0.08,0.5}
\definecolor{MyDarkRed}{rgb}{0.7,0.02,0.02}
\definecolor{MyDarkGreen}{rgb}{0.0,0.7,0.0}

\newcommand{\utokyo}{Department of Astronomy, Graduate School of Science, The University of Tokyo, 7-3-1 Hongo, Bunkyo-ku, Tokyo 113-0033, Japan}
\newcommand{\naoj}{National Astronomical Observatory of Japan, 2-21-1 Osawa, Mitaka, Tokyo 181-8588, Japan}
\newcommand{\nrao}{National Radio Astronomy Observatory, 520 Edgemont Rd, Charlottesville, VA 22903, USA}
\newcommand{\haystack}{Massachusetts Institute of Technology, Haystack Observatory, 99 Millstone Rd, Westford, MA 01886, USA}
\newcommand{\bhi}{Black Hole Initiative, Harvard University, 20 Garden Street, Cambridge, MA 02138, USA}
\newcommand{\ism}{The Institute of Statistical Mathematics, 10-3 Midori-cho, Tachikawa, Tokyo 190-8562, Japan}
\newcommand{\sokendaiism}{Department of Statistical Science, School of Multidisciplinary Sciences, Graduate University for Advanced Studies, 10-3 Midori-cho, Tachikawa, Tokyo 190-8562, Japan}
\newcommand{\sokendainaoj}{Department of Astronomical Science, School of Physical Sciences, Graduate University for Advanced Studies, 2-21-1 Osawa, Mitaka, Tokyo 181-8588, Japan}
\newcommand{\kogakuin}{Division of Liberal Arts, Kogakuin University, 1-24-2 Nishi-Shinjuku, Shinjuku-ku, Tokyo 163-8677, Japan}

\begin{document}
\title{Super-resolution Imaging of the Protoplanetary Disk HD~142527 Using Sparse Modeling}
%
\correspondingauthor{Masayuki Yamaguchi}
\email{masayuki.yamaguchi.astro@gmail.com}
\author{Masayuki Yamaguchi}
\affil{\utokyo}
\affil{\naoj}
\author{Kazunori Akiyama}
\altaffiliation{NRAO Jansky Fellow}
\affil{\naoj}
\affil{\nrao}
\affil{\haystack}
\affil{\bhi}
\author{Takashi Tsukagoshi}
\affil{\naoj}
\author{Takayuki Muto}
\affil{\kogakuin}
\author{Akimasa Kataoka}
\affil{\naoj}
\affil{\sokendainaoj}
\author{Fumie Tazaki}
\affil{\naoj}
\author{Shiro Ikeda}
\affil{\ism}
\affil{\sokendaiism}
\author{Misato Fukagawa}
\affil{\utokyo}
\affil{\naoj}
%
%
\author{Mareki Honma}
\affil{\utokyo}
\affil{\naoj}
\affil{\sokendainaoj}
\author{Ryohei Kawabe}
\affil{\utokyo}
\affil{\naoj}
\affil{\sokendainaoj}
%
%
\begin{abstract}
With an emphasis on improving the fidelity even in super-resolution regimes, new imaging techniques have been intensively developed over the last several years, which may provide substantial improvements to the interferometric observation of protoplanetary disks. In this study, sparse modeling (SpM) is applied for the first time to observational data sets taken by the Atacama Large Millimeter/submillimeter Array (ALMA). The two data sets used in this study were taken independently using different array configurations at Band 7 (330 GHz), targeting the protoplanetary disk around HD 142527; one in the shorter-baseline array configuration ($\sim 430$ m), and the other in the longer-baseline array configuration ($\sim 1570$ m). The image resolutions reconstructed from the two data sets are different by a factor of $\sim3$. We confirm that the previously known disk structures appear on the images produced by both SpM and CLEAN at the standard beam size. The image reconstructed from the shorter-baseline data using the SpM matches that obtained with the longer-baseline data using CLEAN, achieving a super-resolution image from which a structure finer than the beam size can be reproduced. Our results demonstrate that on-going intensive development in the SpM imaging technique is beneficial to imaging with ALMA.
\end{abstract}
\keywords{techniques: high angular resolution --- techniques: image processing --- techniques: interferometric --- ISM: individual objects (HD~142527) --- protoplanetary disks }
%

\section{Introduction}\label{sec:1}
\begin{figure*}[t]
\centering
\includegraphics[width=0.8\textwidth]{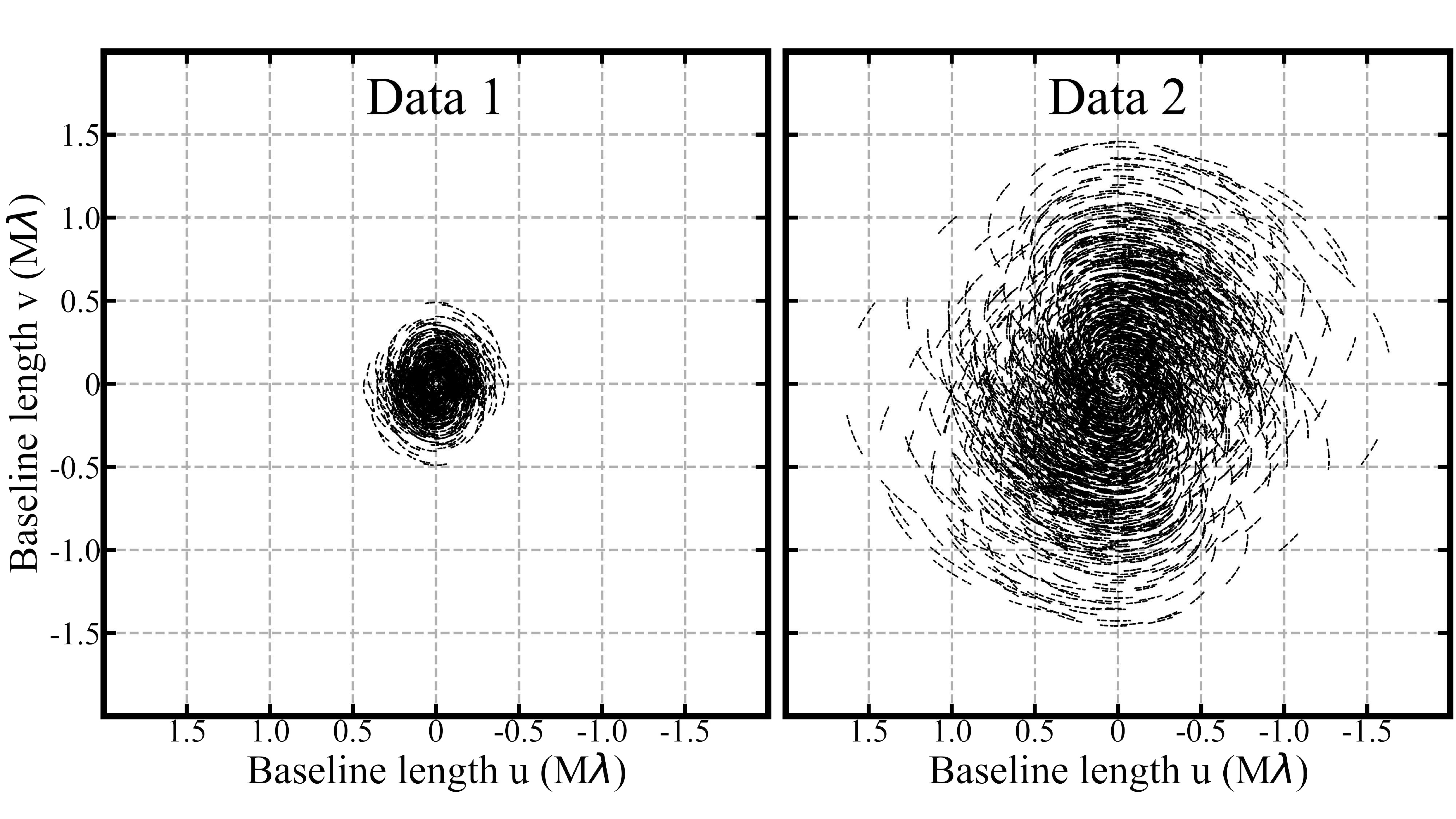}
\caption{$uv$-coverage of two data sets of ALMA observations toward HD~142527 adopted in this work. The left panel shows Data 1 obtained with the more compact array configuration, while the right panel shows Data 2 from the more extended array configuration.}
\label{fig:uv-cov}
\end{figure*}

High-resolution observations are essential in several fields of astronomy because higher angular resolution provides better information concerning the detailed structure of astronomical objects. Interferometry is an effective approach used to obtain images of high angular resolution at radio wavelengths. For example, the Atacama Large Millimeter/submillimeter Array (ALMA) interferometer has revealed small-scale substructures within protoplanetary disks, providing us with valuable insights into planet formation \citep[e.g.,][]{alma2015, Andrews2018}, and ground-based Very Long Baseline Interferometry (VLBI) observations with the Event Horizon Telescope \citep[EHT;][]{doeleman2009} have allowed the unveiling of the compact structure of the magnetized plasma on scales of a few Schwarzschild radii in the vicinity of the supermassive black hole (SMBH) in the nucleus of M87 \citep{eht2019}.

One of the issues related to interferometric observations is that the produced data set is an incomplete set of Fourier components describing an intensity distribution. Since the incomplete set always causes an ``underdetermined problem'' in the radio interferometer equation, the intensity distribution cannot be recovered directly. A means of reconstructing the image is therefore essential.

The CLEAN algorithm  \citep[e.g.,][]{hogbom1974,clark1980, schwab1984,cornwell2008,rau2011} is the most standard image reconstruction algorithm from the data set. The CLEAN algorithm iteratively determines the point source on the image domain that best fits the observed visibilities, starting from an image (i.e., ``dirty image'') made by the inverse transform of the Fourier measurements with all non-observed data set to zero. This process is repeated until some convergence requirement is met. The final image is obtained by convolving the point source model with an idealized CLEAN beam (usually an elliptical Gaussian fitted to a synthesized beam). This algorithm has been widely used for reconstructing images taken at radio interferometers, but recently, a new technique using the sparse modeling (henceforth SpM) approach is proposed to obtain higher resolution images.

SpM is one of the techniques developed for imaging with EHT \citep{honma2014, ikeda2016, akiyama2017a, akiyama2017b, obuchi2017, Kuramochi2018, eht2019d}. It has been proven that this technique achieves a higher fidelity image than the conventional CLEAN algorithm at the angular scale of $30-40~\%$ of the CLEAN beam (i.e., super-resolution) based on mock observations with VLBI \citep[see, e.g.][and references therein]{chael2016, chael2018, akiyama2017a, akiyama2017b, Kuramochi2018}. Here, ``fidelity'' is a measure of how brightness distribution on a reconstructed image is faithfully restored to an expected astronomical object. As an example, the image fidelity is quantitatively measured by some popular metrics such as the normalized root mean square error (NRMSE; see Section \ref{sec:3_nrmse} for the definition).

SpM has been applied for the imaging of the shadow of a black hole using the EHT \citep{eht2019}, VLBI observations of high-dynamic range images of relativistic jets \citep[][]{tazaki2018}, and Very Large Array (VLA) observations of the stellar photosphere \citep{matthews2018}. SpM has also been used to provide high-quality reconstructions of full polarization imaging on a mock observation with VLBI \citep{akiyama2017b} and mathematically similar Rotation Measure synthesis by using simulation data \citep{akiyama2018}. However, it has not yet been applied to ALMA observational data, and it is therefore unknown whether SpM imaging is useful for the reconstruction of super-resolution images based on actual observational data. Although SpM is as yet under development, current techniques already have the potential to improve ALMA images significantly.

In this study, for the first time, SpM imaging is applied to an ALMA observational data set of the protoplanetary disk around HD~142527. The target object hosts one of the most well-studied transition disks. It is a binary system at a distance of 156~$\pm$~7.5~pc \citep{gaia2016b,gaia2016a}. The primary star is a Herbig Ae/Be with spectral type F6 III, having a mass of 2.2~$\mbox{M}_{\odot}$ while the secondary has a mass of 0.1-0.4~$\mbox{M}_{\odot}$ \citep{Verhoeff2011, Biller2012}. Several observations of the object have so far been carried out with ALMA. The results show that the brightness distribution is strongly lopsided and that the radius of the cavity is $\sim 150 $ au \citep{Casassus2013, fukagawa2013, boehler2017, Ohashi2018, soon2019}. An observational data set covering several different angular resolutions therefore exists in the same frequency band. It is possible to evaluate the performance of SpM imaging by comparing the resulting images with those derived using the CLEAN algorithm.

In this study, two sets of ALMA archive data at Band 7 ($\sim$330GHz) were used to investigate the protoplanetary disk around HD 142527, one of which was taken with the compact array configuration, and the other with the extended array configuration. The maximum baseline lengths between the two data sets are different by a factor of $\sim3-4$. Images were constructed from the data sets using the SpM and the popular multi-scale Cotton-Schwab CLEAN algorithm \citep[hereafter MS-CLEAN; ][]{cornwell2008,rau2011} to evaluate whether the previously seen disk structures appear on the images made by both the SpM and MS-CLEAN at the angular resolutions taken by each data set. The fidelity of the SpM image was also compared with the image from MS-CLEAN by changing the angular resolution for both images. The detailed outer disk structure of the SpM image seen in the super-resolution regime is also discussed.

This study provides the first opportunity to evaluate the performance of the SpM imaging using real observational data at different angular resolutions. The paper is organized as follows. The observations, calibrations, and imaging procedures are introduced in Section \ref{sec:2}, the images from both data sets are then evaluated based on a general image fidelity metric in Section \ref{sec:3}. The disk substructure inferred from the SpM image at the super-resolution regime and the remaining technical issues are discussed in Section \ref{sec:4}. The conclusion of this study is presented in Section \ref{sec:5}.

\section{Data Reductions and Imaging}\label{sec:2}
\subsection{ALMA Data Set used for Imaging}\label{sec:2.0}
Two data sets of ALMA observations of the protoplanetary disk around HD~142527 at a frequency of $\sim$ 330 GHz (ALMA Band 7) are used in our investigation. One uses a compact antenna configuration, and the other uses extended. Figure \ref{fig:uv-cov} shows the $uv$-coverage of the two data sets. The maximum extensions of the baseline lengths differ by a factor of $\sim$3-4, leading to the same difference in the size of the synthesized beams.  The calibrations used for each data set are summarized below.

The data set obtained with the compact array configuration with a maximum baseline length of 430~m was labeled as {\it Data 1}. Data 1 were obtained as part of the project $2015.1.00425$.S, which has already been published in \citet{kataoka2016}. The corresponding observations were carried out on March 11, 2015, at 343 GHz (0.87 mm) to detect the full polarization of the continuum emission. The observing array consisted of thirty-eight 12~m antennas. The data were taken over a total bandwidth of 8~GHz, consisting of four 2~GHz spectral windows centered at 336, 338, 348, and 350~GHz. The on-source time of HD~142527 was 1.2~h, and observations were carried out for 3.4~h in total.

The data set obtained with the extended array with a maximum baseline length of 1570~m was labeled as {\it Data 2}. Data 2 were obtained as part of the project $2012.1.00631$.S, which was carried out on July 17, 2015, at 322 GHz (0.93 mm) with the correlator configuration for dual-polarization. The observations made use of forty 12~m antennas. The total observing time was 22.5 h with an on-source time of 1.9 h. The total bandwidth of 4.8 GHz was separated into two 0.5~GHz spectral windows centered at 314 and 329~GHz and two 1.9~GHz spectral windows centered at 315 and 328~GHz.

\subsection{Data Reduction and Imaging with MS-CLEAN}\label{sec:2.2}

Data 1 were calibrated using version 4.7.2 of the Common Astronomy Software Applications package \citep[CASA;][]{mcmullin2007}, in the same manner as \citet{kataoka2016}. The initial calibration was performed using the ALMA pipeline. In the pipeline, the complex gains and bandpass were calibrated with J1604-4441 and J1427-4206, respectively, while the instrumental polarization was calibrated with J1512-0905. To improve the fidelity of the image, we performed the self-calibration technique for the corrected data. First, we constructed Stokes I model of HD 142527 using MS-CLEAN performed with scale parameters of [0, 0.3, 0.9] asec (``asec'' is an abbreviation for ``arcsecond'') by adopting Briggs weighting of robust parameter 0.5. Next, using the MS-CLEAN model, we performed iterative self-calibration of the visibility phase (calmode = p). The interval time used to solve the complex gain varied from 420 to 30 s. The resultant image (= MS-CLEAN model convolved with CLEAN beam $+$ residual map) after self-calibration provided the beam size of $0.51 \times 0.44$~asec with position angle (P.A.) of 58.7$^\circ$. The RMS noise level of the resultant image was $0.32~ \rm mJy~beam^{-1}$.

Data 2 were calibrated in the same manner as Data 1. The data were initially calibrated with the same version of the ALMA pipeline. In the pipeline, Pallas, Ceres, and J1427-4206 were used for the flux calibration, while the complex gains and bandpass were calibrated with J1604-4228 and J1517-2422, respectively. The corrected data were imaged using MS-CLEAN performed with scale parameters of [0, 0.3, 0.9] asec by adopting Briggs weighting of robust parameters 0.5. The corrected data were then further calibrated iteratively with MS-CLEAN and self-calibration in phase (calmode = p). The interval of time used to solve the complex gain varied from 360 to 50 s. The resultant image (= MS-CLEAN model convolved with CLEAN beam $+$ residual map) after self-calibration provided the beam size of $0.20 \times 0.14$~asec at P.A. of 78.1$^\circ$. The RMS noise level of the resultant image was $0.07~ \rm mJy~beam^{-1}$.

We note that the ratio of the central frequencies of the two data sets is $\sim 0.9$. In the (sub) millimeter continuum emissions of protoplanetary disks, the ratio may cause a difference of $\sim 10-20~\%$ in the source intensity based on the typical value of the spectral index ($\alpha_{\rm mm} \sim 2-3$ given by $F_{\nu} \propto \nu^{\alpha_{mm}}$), such as that typically seen in the \citep{beckwith1991,mannings1994,andrews2005,andrews2007}. With a typical flux calibration error of up to $10~\%$ in ALMA observations \citep{mathys2013, andreani2015}, the source intensity of the two data sets may differ by a total of $\lesssim 30~\%$. To check the difference in intensity between the two data sets, the total fluxes are estimated by measuring the maximum value of the visibility amplitude. The resultant total fluxes of Data 1 and 2 are derived to be 3.3 Jy and 3.2 Jy, respectively, indicating a total difference of $\sim3~\%$. The two data sets therefore satisfy the assumptions.

\subsection{Imaging with Sparse Modeling}\label{sec:2.3}


\begin{figure*}[h]
\centering
\includegraphics[width=0.95\textwidth]{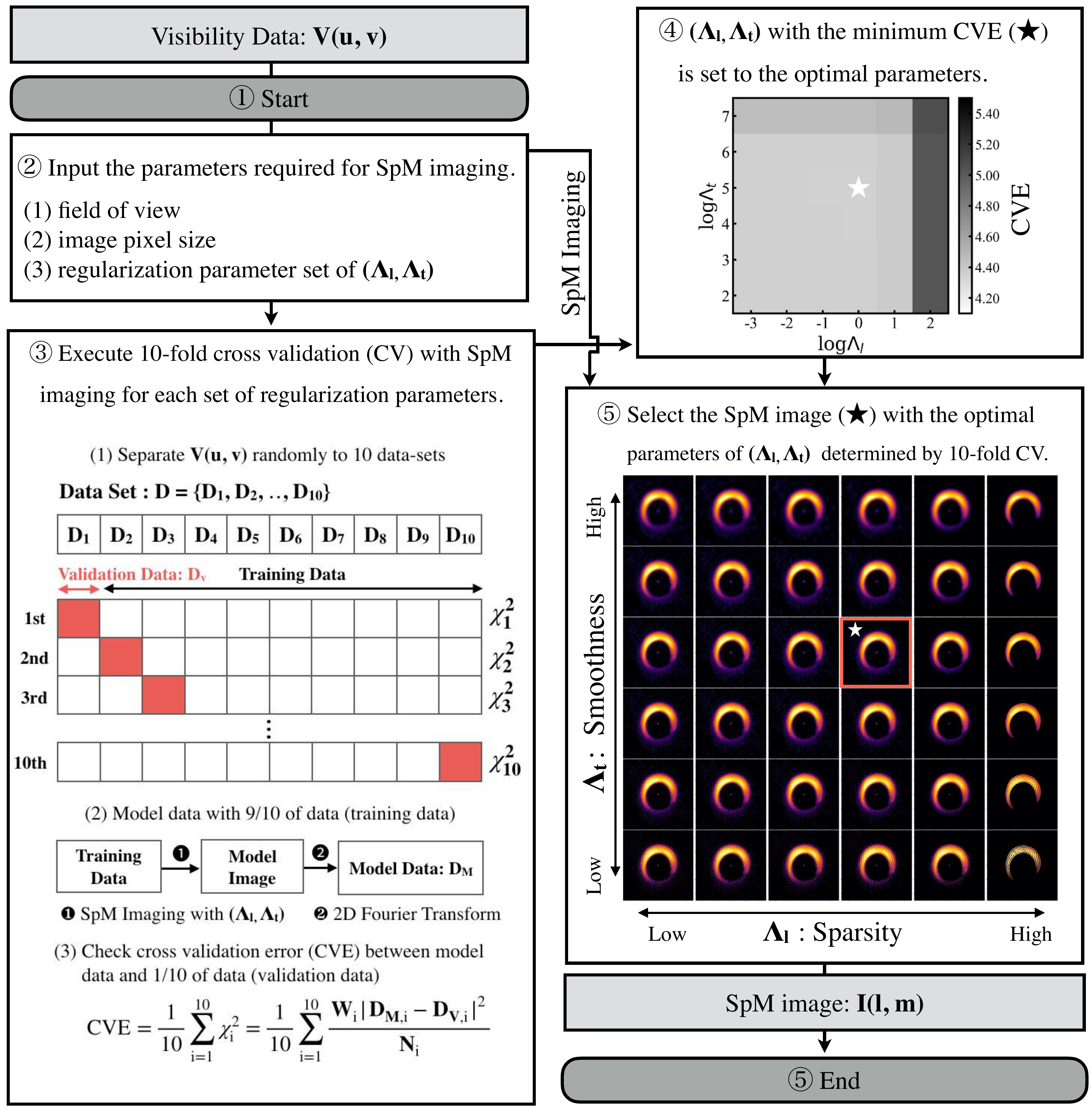}
\caption{Flowchart of the image processing with SpM. The SpM image is automatically generated through 10-fold cross validation (CV) after inputting a user-specified parameter set. Step 1: we prepare the visibility data set (here, the self-calibrated visibilities of data set is used). Step 2: the user-specified parameters (i.e., field of view, image pixels size, and regularization parameter set of $(\Lambda_{l}$, $\Lambda_{t})$) is set. Step 3: 10-fold cross validations with SpM imaging for each set of regularization parameters are executed, providing the results of cross validation error (CVE). In the equation of CVE, N is the total training data, and W ($=1/\sigma^{2}$) is the weight for the visibilities.  Step 4: $(\Lambda_{l}$, $\Lambda_{t})$ with the minimum CVE is set to the optimal regularization parameters. Step 5: SpM imaging with full data set is executed, and the SpM image with the optimal regularization parameters is finally selected.}
\label{fig:spm_flowchart}
\end{figure*}

We used the self-calibrated visibilities of both data sets to reconstruct the images with SpM utilizing $\ell _1$+TSV regularization \citep[][]{Kuramochi2018}. This latest imaging technique utilizes two convex regularization functions of the brightness distribution: $\ell_1$-norm and total squared variation (TSV). These regularizers adjust sparsity and smoothness in the brightness distribution, respectively. This technique can be used to achieve resolutions as high as $\sim30-40 \%$ of the angular resolution while maintaining image fidelity \citep{Kuramochi2018}. The imaging equation is formulated as following.

\begin{eqnarray}\label{spm_eq}
{\bf I} & = & \argmin_{\bf I} \Big(||{\bf W}({\bf V} - {\bf FI})||^2_2 + \Lambda_l \sum_i \sum_j |\rm I_{i,j}| \nonumber \\
& + & \Lambda_t \sum_i \sum_j \big(|\rm I_{i+1,j} - \rm I_{i,j}|^2 + |\rm I_{i,j+1} - \rm I_{i,j}|^2\big)\Big),
\end{eqnarray}
\noindent
where ${{\bf I} = \{\rm I_{i,j}\}}$ is the two dimensional image reconstruction to be solved, the element in row $i$ and column $j$ is represented by $\rm I_{i,j}$, ${\bf V}$ is the observed visibility (i.e., the self-calibrated visibilities), ${\bf F}$ is the Fourier matrix, and ${\bf W}=\{\delta _{ij} /\sigma^{2}_{ij}\}$ is a diagonal matrix normalizing the residual visibility $({\bf V} - {\bf FI})$ on the first term. $\sigma _{ij}$ is the observational error of each data point, and $\delta _{ij}$ is the Kronecker delta. The first term is the sum of the squared residuals between the observational data and the model, namely the traditional $\chi^2$-term that represents how well the reconstructed image reproduces the observational data. The second term, $\ell_1$-norm, adjusts the sparsity of the brightness distribution, while the last term handles the sparsity of the TSV in the gradient domain which effectively controls the smoothness of the brightness distribution. The balance of these two regularization terms is controlled by two positive variables $\Lambda_{l}$ and $\Lambda_{t}$.

A procedure of generating an SpM image from visibility data is illustrated in the flowchart in Figure \ref{fig:spm_flowchart}. The parameters first required for SpM imaging are (1) field of view, (2) pixel size of the reconstructed image, and (3) two regularization parameters for regularization functions, namely $\ell_1$-norm and TSV \citep{Kuramochi2018}.

A pixel size of $0.05\: \times\: 0.05$~asec was used for Data 1 and $0.025\: \times\: 0.025$~asec for Data 2, which were both at least 7 times smaller than the synthesized beam of MS-CLEAN, and $5.0 \times 5.0 $~asec for the field of view, which is large enough to cover the entire region where the continuum emission has been detected. Note that this pixel size does not significantly affect the resultant images as it is small enough to trace the structure on the spatial scales constrained by the longest-baseline visibilities, because we utilize the TSV which supports multi-resolution reconstruction by regularizing the gradient function of the image. The TSV can reconstruct an edge-smoothed image \citep{Kuramochi2018}.

we adopt $16\times9$ sets of regularization parameters for Data 1, consisting of ($1\times10^{2}, 2\times10^{2}, ..., 9\times10^{2}$) for $\Lambda_{l}$ and ($3\times10^{4}, 4\times10^{4}, ..., 9\times10^{5}$) for $\Lambda_{t}$, and $6\times6$ sets of regularization parameters for Data 2, consisting of $(10^{2}, 10^{3}, ..., 10^{7})$ for $\Lambda_{l}$ and $(10^{-3}, 10^{-2}, ..., 10^{2})$ for $\Lambda_{t}$. The optimal parameter set of ($\Lambda_{l}$, $\Lambda_{t}$) was determined by 10-fold cross validation (CV) \citep[see][for details]{akiyama2017a,akiyama2017b} that evaluates the parameter sets and chooses a parameter set providing the optimal goodness-of-fit for given uncertainties. The cross validation error (CVE) of these parameter sets was then evaluated using 10-fold CV, and the parameter set and corresponding image were selected, minimizing the CVE for each spectral window of each set of data.

For the self-calibrated visibilities, the imaging equation (\ref{spm_eq}) becomes a convex optimization, guaranteeing the convergence to a unique solution regardless of the initial conditions \citep[e.g.,][]{akiyama2017b,Kuramochi2018}. The fast iterative shrinking threshing algorithm \citep[FISTA;][]{beck2009a,beck2009b} is a popular and efficient algorithm for solving optimization; therefore a monotonic variant of FISTA (MFISTA) was used, that is especially designed for the regularization of $\ell_1$-norm with another convex function \citep[see][for details]{akiyama2017b}. Prior to imaging, the self-calibrated visibilities were gridded using cell-averaging \citep[e.g., see][]{thompson2017}. Images were reconstructed for both data sets at each spectral window to evaluate the noise levels in the reconstructed images, and also to minimize the potential effects caused by frequency-dependent residual gains in the visibility amplitudes. As a result, four images were reconstructed for each data set and then averaged into a final image.

It is worth noting that because the self-calibrated visibilities that were iteratively calibrated with MS-CLEAN imaging and self-calibrations in phase were used, the SpM images will be affected by residual gains in the visibility amplitude, as well as the residual phase errors induced by the use of self-calibration with MS-CLEAN, which may cause additional errors and artifacts on the reconstructed images. The effects of these residual complex gains are discussed in Section \ref{sec:4_noise} in detail, although it will not significantly affect the main results of this paper described in Section \ref{sec:3}.

\begin{figure*}[t]
\centering
\includegraphics[width=0.9\textwidth]{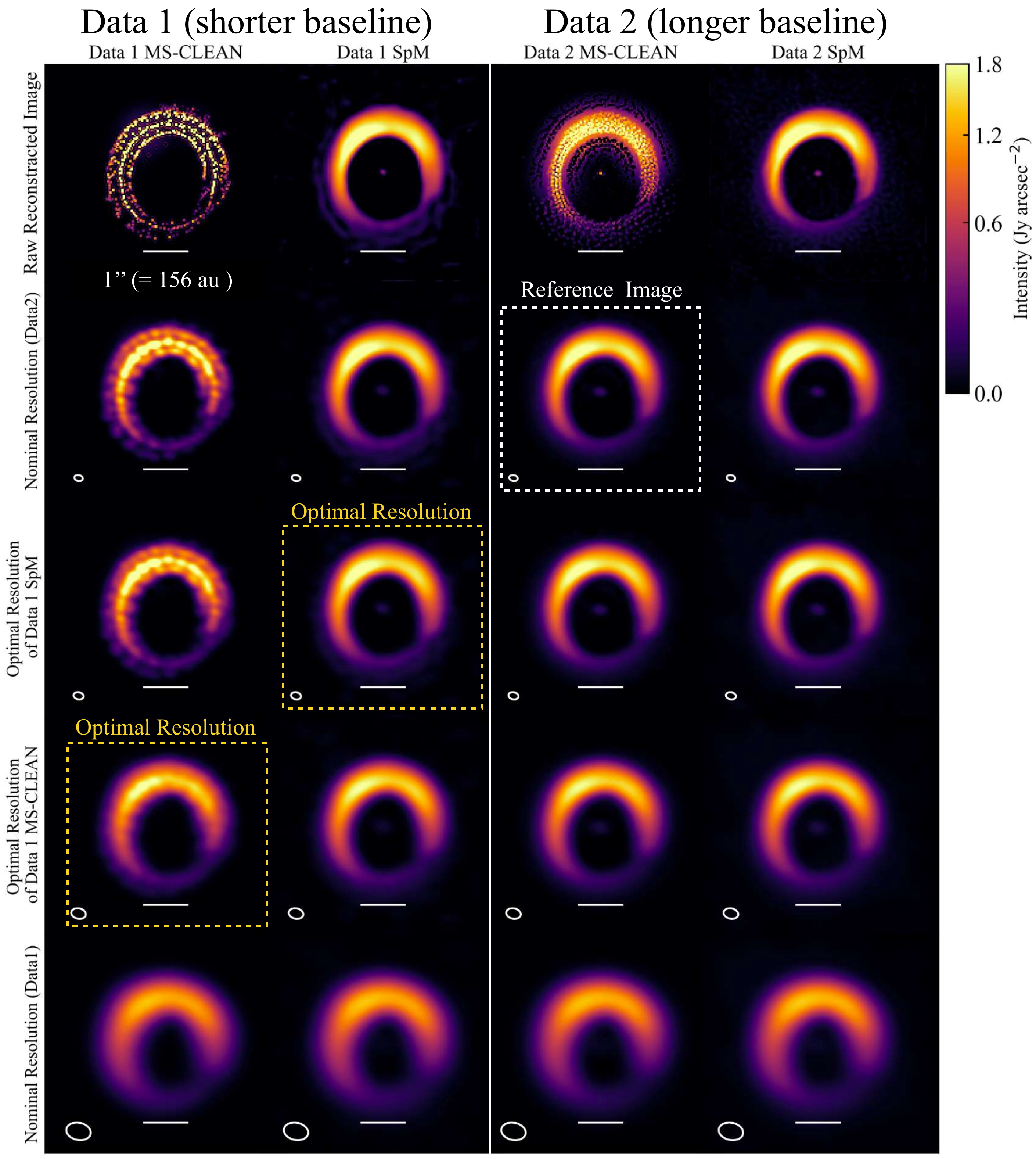}
\caption{Images of HD~142527 constructed with two data sets from ALMA observations taken at 322 and 343 GHz as reconstructed using SpM and MS-CLEAN. The same color scale given by a power law with a scaling exponent of 0.6 and field of view of 5.0 $\times$ 5.0 asec are adopted for all images.
The left two columns show images from Data 1 reconstructed with SpM and MS-CLEAN, respectively, while the right two columns show images from Data 2.
The raw reconstructed images or those restored with an elliptical Gaussian beam are given for each row of images, for which the FWHM shape is shown in a white ellipse in each panel.
The axial ratio and P.A. of the beam are fixed to those of the synthesized beam size of Data 2 for Briggs weighting with a robust parameter of 0.5 adopted for use with MS-CLEAN. 
(Top panels): The raw reconstructed images without any Gaussian convolution.
(2nd panel): Reconstructed images convolved with the nominal resolution of Data 2.
(3rd panel): Reconstructed images convolved with the optimal resolution of Data 1 for SpM, respectively, determined by NRMSE analysis (see Section \ref{sec:3_nrmse} and Figure \ref{fig:nrmse}).
(4th panel): Reconstructed images convolved with the optimal resolution of Data 1 for MS-CLEAN, determined by NRMSE analysis.
(5th panels): Reconstructed images convolved with the nominal resolution of Data 1}

\label{fig:images}
\end{figure*}

\section{Results}\label{sec:3}
\subsection{Images at Different Angular Resolutions}
The images reconstructed by SpM and MS-CLEAN are compared and evaluated to ascertain whether the previously known disk structures appear on both images. The image fidelity of SpM in comparison with MS-CLEAN is also examined with regards to the super-resolution regime. Figure~\ref{fig:images} shows the reconstructed images from the two data sets of the ALMA observations at 322 and 343 GHz using SpM and MS-CLEAN. The images are either not convolved (top panels) or convolved (lower panels) with different sizes of elliptical Gaussian beams. Nominal resolutions are defined as synthesized beams of the MS-CLEAN images for Data 1 and 2. Meanwhile, in both the SpM and MS-CLEAN images, optimal resolutions are determined using normalized root mean square error (NRMSE) analysis by regarding the Data 2 MS-CLEAN image as the reference image (see Section \ref{sec:3_nrmse}). For the NRMSE analysis, the beam and P.A. of the images has to be matched to that of the reference image. In this analysis, the nominal resolution of Data 1 is therefore modified to $0.57\times0.40$ asec with a P.A. of $78.1^{\circ}$, which is determined to have the same solid angle of the beam as the original one.

We define that the beam size ratio for the nominal resolution of Data 1 (labeled by {\it R}) has a range of $0~\% \leq \rm R \leq 100~\%$. The lowest angular resolution shown in the bottom panels of Figure~\ref{fig:images} corresponds to the nominal resolution of Data 1 ($0.57\times0.40$ asec, R = $100~\%$). It is apparent that the source intensity is consistent within $\lesssim 6~\%$ between the four images, which may be accounted for by the differences in the central frequencies and the flux calibration errors between Data 1 and 2. All the four images show a lopsided structure in the outer disk and the thermal dust emission that is brighter in the northeastern side, as seen in previous studies \citep{Casassus2013,fukagawa2013, Casassus2015, boehler2017, Ohashi2018, soon2019}.

The disk structures in the SpM and MS-CLEAN images start to deviate at the optimal resolution of Data 1 MS-CLEAN ($0.34 \times 0.24$ asec, R = $60~\%$). Blobby structures appear in the MS-CLEAN image, while those are blurred. The blobby structures become more apparent in the MS-CLEAN image at the optimal resolution of Data 1 SpM ($0.23 \times 0.16$ asec, R = $41~\%$). On the other hand, the Data 1 SpM image is still consistent with the Data 2 MS-CLEAN image at these resolutions, even in a super-resolution regime (i.e., the nominal resolution of MS-CLEAN for Data 2, R = $35~\%$). This fact is consistent with previous works based on imaging simulations \citep[e.g.,][]{chael2016,akiyama2017a,akiyama2017b, Kuramochi2018}.

It is worth noting that the differences between SpM and MS-CLEAN are also remarkable in the raw reconstructed images which present the original images before Gaussian convolution (top panels of Figure~\ref{fig:images}, R = $0~\%$). The raw SpM images from the two data sets with different array configurations consistently show a smooth distribution of brightness in the outer disk. In comparison, the outer disk in the raw MS-CLEAN images, in which the map of clean components is shown, consists of more compact point-like features, which are not consistent with each other or with any of the other images, suggesting that these features can be presumed artificial.

\subsection{Fidelity at Multi-resolution}\label{sec:3_nrmse}
For more quantitative analysis, we evaluate the normalized root mean square error (NRMSE) between the images with different angular resolutions \citep{chael2016,akiyama2017a,Kuramochi2018}. The NRMSE is defined as:
\begin{equation}
\label{eq:nrmse_im}
{\rm NRMSE}(\mathbf{I}, \mathbf{K})_{\rm image} =\sqrt{\frac{\sum_{i}\sum_{j}|\rm I_{i,j}-\rm K_{i,j}|}{\sum_{i}\sum_{j}|\rm K_{i,j}|}}.
\end{equation}
\noindent
where $\mathbf{I}=\{\rm I_{i,j}\}$ is the input image and $\mathbf{K}=\{\rm K_{i,j}\}$ is the reference image. The NRMSE is calculated by changing a beam size (i.e., a spatial resolution). The beam size providing the minimum NRMSE may be considered as an optimal resolution \citep{chael2016}. Previous studies \citep[e.g.,][]{akiyama2017a,Kuramochi2018} suggest that the NRMSE is often dominated by errors in excessively bright pixels, and therefore may not represent the fidelity of some other properties such as the smoothness of the image and the size of the emission region. Hence, the NRMSE of the gradient-domain brightness distribution was also evaluated using the Prewitt filter \citep[][]{Kuramochi2018}. The metric for the fidelity of the image is evaluated by taking the gradients of the image, given by
\begin{equation}
\label{eq:nrmse_grad}
| \nabla \rm I ( x , y ) | = \sqrt { \left| \frac { \partial I } { \partial x } \right| ^ { 2 } + \left| \frac { \partial I } { \partial y } \right| ^ { 2 } }
\end{equation}
as a continuous distribution of the brightness. Errors in the image gradients can be evaluated using equation (\ref{eq:nrmse_im}),
\begin{equation}
{\rm NRMSE } _ {\rm grad} ( \mathbf{I} , \mathbf{K}) \equiv {\rm NRMSE} _ {\rm image}( \nabla \mathbf{I} , \nabla \mathbf{K} )
\end{equation}

To calculate NRMSE, it is necessary to choose a reference image. The risk of intrinsic bias becomes lower when the reference image is chosen to be a method that is well known to and trusted by the community. The Data 2 MS-CLEAN at its nominal resolution is therefore the best choice as the reference image. With this reference image, the NRMSEs of the beam-convolved MS-CLEAN and SpM images of Data 1 are evaluated.

\begin{figure*}[ht]
\centering
\includegraphics[width=1\textwidth]{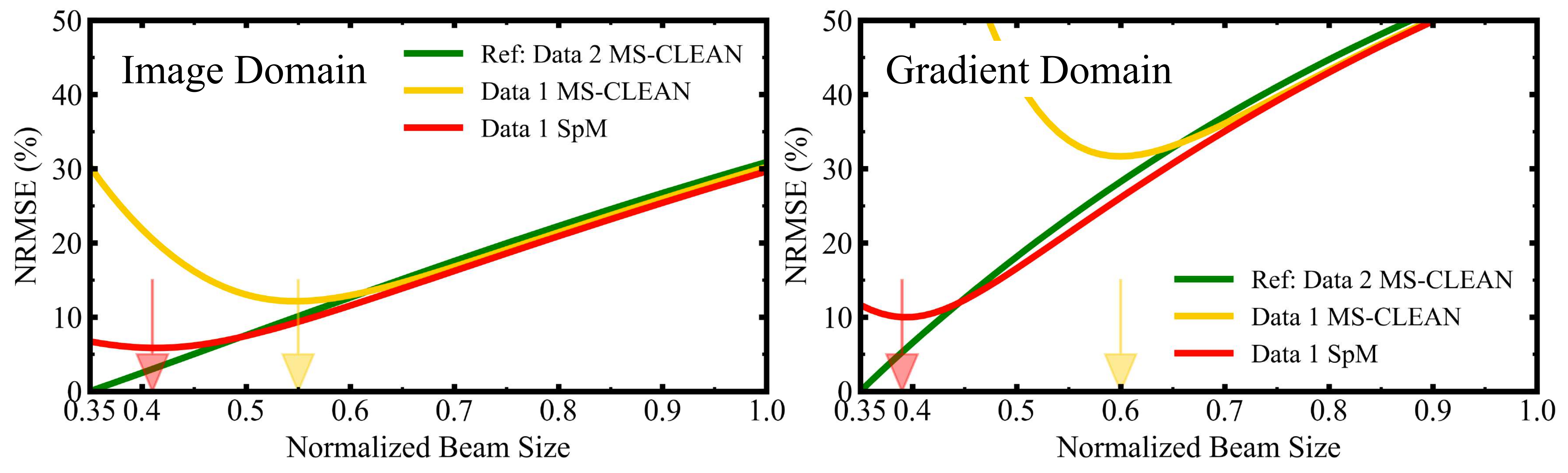}
\caption{NRMSEs of the four reconstructed images as a function of the normalized beam size, on the image domain (the left panel), and the gradient domain (the right panel). The Data 2 MS-CLEAN image is adopted as the reference image (green line) for the NRMSE analysis in both panels. Each image is convolved with an elliptical beam size, with the same axial ratio and P.A. as the nominal resolution of Data 2. The nominal resolution of Data 1 normalizes the size of the convolving beam shown in the horizontal axis. The nominal resolution of Data 2 indicates $35~\%$ ($= 20\times 14$ mas) of Data 1.}
\label{fig:nrmse}
\end{figure*}

Figure \ref{fig:nrmse} shows the results of NRMSE analysis on both the image and gradient domains. Each image is convolved with an elliptical beam size, whose axial ratio and P.A. are the same as that of the Data 2 MS-CLEAN image. The nominal resolution of Data 1 normalizes the convolving beam size. The solid green line shows the ideal NRMSE curves between the non-convolved and the convolved reference images, effectively quantifying the best-case scenario in which the difference from the original input is due to a loss of resolution. The other lines show the NRMSEs between the reference image and the other three images. 

Table \ref{tab:beamsize} summarizes the optimal resolutions as determined by the NRMSE. For each image, the worst optimal resolution was selected from the results of two NRMSE analyses (for the image and gradient domains) and defined as the optimal resolution. In the 3rd and 4th rows of Figure \ref{fig:images}, we show all four images convolved with the optimal resolutions of the SpM and MS-CLEAN images, respectively. In the following paragraphs, we describe in more detail the characteristics of each curve (image and gradient domain) and their relation to the corresponding images.

The Data 1 MS-CLEAN image achieves the optimal resolution at R = $60~\%$ and sharply increases the NRMSEs at finer resolutions. The compact artifacts created in MS-CLEAN dominate the deviations from the other lines for NRMSE. The Data 1 MS-CLEAN image at its optimal resolution (4th row of Figure \ref{fig:images}) already shows blobby-like structures, which can be attributed to the underlying assumption used in the production of the image.

In general, the original image of MS-CLEAN is composed of an ensemble of point sources and Gaussian sources with appropriate sizes, i.e., CLEAN components (top row of Figure \ref{fig:images}). The final CLEAN image is reconstructed by convolving the CLEAN components with an idealized CLEAN beam corresponding to the nominal resolution of Data 1. With a finer spatial resolution than the nominal resolution, the MS-CLEAN image gets close to the CLEAN components itself, making the image blobby. Since NRMSE is more weighted at brighter pixels than lower-intensity skirts of the emission, the optimal resolution would be slightly finer than the nominal resolution, causing the blobby structures. These results are consistent with previous work on imaging simulations \citep{chael2016,akiyama2017a,akiyama2017b, Kuramochi2018}.

In contrast, the Data 1 SpM image follows the curves of the reference image until at R = $45-50~\%$ in both the image and in the gradient domains. The optimal resolution reaches R = $41~\%$, which is better than that of the Data 1 MS-CLEAN. At resolutions finer than R = $41~\%$, the NRMSE of the Data 1 SpM shows flat curves until at the nominal resolution of Data 2 (R = $35~\%$), with typical ranges of $5-10~\%$ in the image domain and $10-15~\%$ in the gradient domain. Although this resolution is better than the nominal resolution for Data 1 by a factor of $\sim3$ (i.e., a super-resolution), the Data 1 SpM image is consistent with the Data 2 MS-CLEAN at the same resolution, while the Data 1 MS-CLEAN image is significantly affected by compact structures that can be attributed to underlying assumption used in the production of these images.

\begin{table*}[ht]
\caption{Optimal Resolutions determined by NRMSE analysis} 
\label{tab:beamsize}
\begin{tabularx}{\linewidth}{lrrr}
\toprule
Angular Resolution & SpM (Data 1) &  MS-CLEAN (Data 1) & Reference: MS-CLEAN (Data 2) \\
\hline
Nominal Resolution (mas) & NaN &  57 $\times$  40~(100~$\%$) & 20 $\times$  14~(35~$\%$) \\
Optimal Resolution on Image Domain (mas) & 23 $\times$  16 ~(41~$\%$) & 31 $\times$  22~(55~$\%$) & NaN \\
Optimal Resolution on Gradient Domain (mas) & 22 $\times$ 15~(39~$\%$) & 34 $\times$ 24~(60~$\%$) & NaN\\
\bottomrule
\end{tabularx}
\tablecomments{The axial ratio and the P.A. of the beams for Data 1 are fixed to those of the synthesized beam size of the MS-CLEAN image for Data 2. The percentages indicate the beam size ratio for the nominal resolution of Data 1. ``mas'' is an abbreviation for ``milliarcsecond''}
\end{table*}



\subsection{Evaluation of Radial and Azimuthal Structure of the Outer Disk}\label{sec:paprofile}
The NRMSE analysis described in Section \ref{sec:3_nrmse} evaluates the image fidelity by compressing two-dimensional information into a single value. In this section, we investigate source-specific quantities to evaluate the reconstructed images.

\begin{figure}[t]
\centering
\includegraphics[width=0.48 \textwidth]{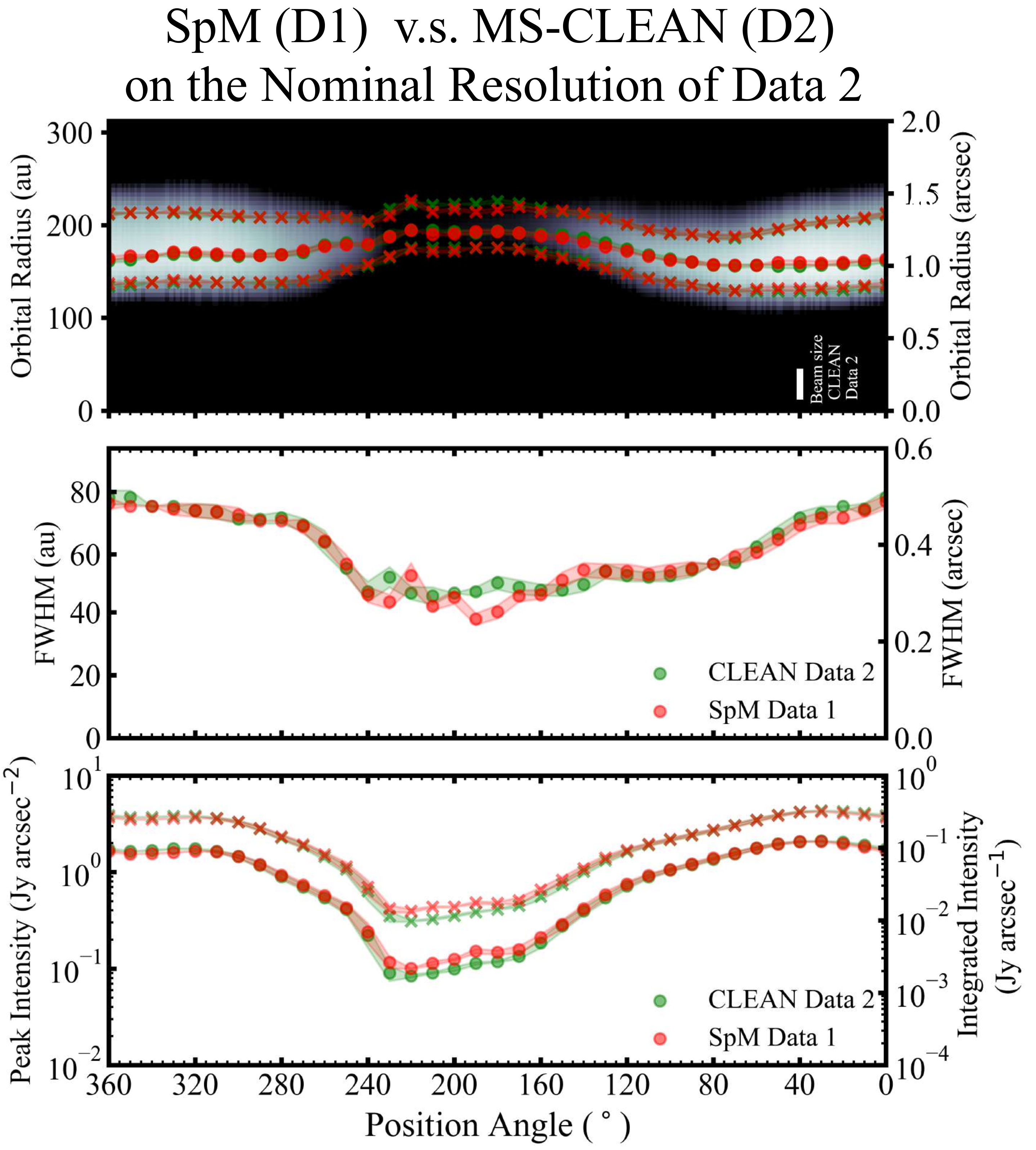}
\caption{The position angle (P.A.) profile for the Data 1 SpM (red color) and the Data 2 MS-CLEAN (green color) images convolved with the Data 2 MS-CLEAN nominal resolution. The top panel shows the radial location of the peak (circle points) and its $50\%$ (i.e., FWHM; cross points) of the intensity distribution of the outer disk overlaid by the Data 2 MS-CLEAN image. The middle panel shows the radial FWHM sizes of the intensity distribution for the outer disk. Bottom panel shows the peak (circle points) and radially integrated (cross points) intensities.}
\label{fig:pa_profile}
\end{figure}

Interesting metrics would be based on the radial and azimuthal structure of the outer disk, with the origin set to the location of the central star. Considering that each radial structure would form a Gaussian distribution, Figure \ref{fig:pa_profile} shows the peak and the FWHM of the radial surface brightness profiles at position angles for the Data 1 SpM and Data 2 MS-CLEAN images convolved with the nominal resolution of Data 2. The data points of the physical parameters of the outer disk, each within $10^\circ$ along the P.A. are thus collected, deriving the mean values as a measurement set on the P.A. profiles. The shaded regions in Figure \ref{fig:pa_profile} indicate the standard deviations $\sigma_{i}$ derived by the calculation. The total flux of the Data 1 SpM image is scaled to that of the Data 2 MS-CLEAN image to minimize the effects of errors in the flux calibration. 

Table \ref{tab:pa_residuals} summarizes the residual statistics for the physical parameters of the P.A. profiles, as shown in Figure \ref{fig:pa_profile}. The residual parameters are subtracted from the two images on each P.A. profile. The radial locations of the peak and its halves (outer/inner half peak) and the radial FWHMs are mostly consistent within nearly 10$\%$ of the nominal angular resolution of Data 2, which is close to the pixel size of the image. The peak and integrated intensity are also consistent; the mean and standard deviation are comparable to the noise levels on the residual maps estimated in Section \ref{sec:4_noise}. Therefore, Figure \ref{fig:pa_profile} indicates that each profile is in good agreement in terms of the radial and azimuthal structure of the outer disk. 

\begin{table}[ht]
\caption{Residual Statistics of P.A. profiles}
\label{tab:pa_residuals}
\begin{tabularx}{\linewidth}{lccccccccc}
\toprule
& \multicolumn{3}{c}{SpM (D1) - MS-CLEAN (D2)}\\
\cmidrule(lr){2-4} \cmidrule(lr){5-7} \cmidrule(lr){8-10}
Quantities (unit)              & Mean     & Std     & Abs.Max   \\ \hline
Radial Peak (mas)              & 1.14     & 14.82    & 27.00    \\
Outer Half Peak (mas)          & 1.38     & 12.58    & 24.00    \\
Inner Half Peak (mas)          & -5.92    & 22.30    & 27.00    \\
FWHM (mas)                     & -7.30    & 21.93    & 36.00    \\
Peak I $\mathrm{(mJy~asec}^{-2})$   & -2.77    & 49.37    & 46.23  \\
Integr. I $\mathrm{(mJy~asec}^{-1})$ & -1.20    & 8.05     & 5.51 \\
\bottomrule
\end{tabularx}
\end{table}


\section{Discussion}\label{sec:4}
\subsection{Investigations of Raw Reconstructed Images}\label{sec:rawspmim}
In Section \ref{sec:3}, we have demonstrated that the application of SpM to Data 1 is possible to provide a high fidelity image in the super-resolution regime. Following the successful experimental application of the SpM, in this section, we discuss the SpM image reconstructed from especially Data 2.

We first compare the SpM and MS-CLEAN images of Data 2 at its nominal resolution. As a result, we find that they are in good agreement at the nominal resolution in terms of the radial and azimuthal structures of the emission, as shown in Appendix.\ref{appendix_a}.

We next focus on the raw SpM image. Figure \ref{fig:images} shows that the raw SpM image from Data 2 is similar to the image that has undergone post-imaging beam convolution at the nominal resolution of Data 2. \cite{Kuramochi2018} pointed out that the raw reconstructed image can keep a high fidelity and thereby the traditional method of the Gaussian convolution with a restoring beam in interferometric imaging would no longer be required for the $\ell _1$+TSV regularization. It is thus instructive to investigate whether the substructures of the emission are seen in the raw reconstructed image of SpM. In the following subsections, we first introduce the noise term of the raw reconstructed image of SpM to evaluate the detection threshold, and the substructures seen in the image are discussed.

In addition, it should be worth showing the NRMSE analysis when the raw SpM image is adopted as the reference. We show this analysis in Appendix.\ref{appendix_b}.

\subsubsection{Noise Terms in Reconstructed Images}\label{sec:4_noise}
\begin{figure}[t]
\centering
\includegraphics[width=0.45\textwidth]{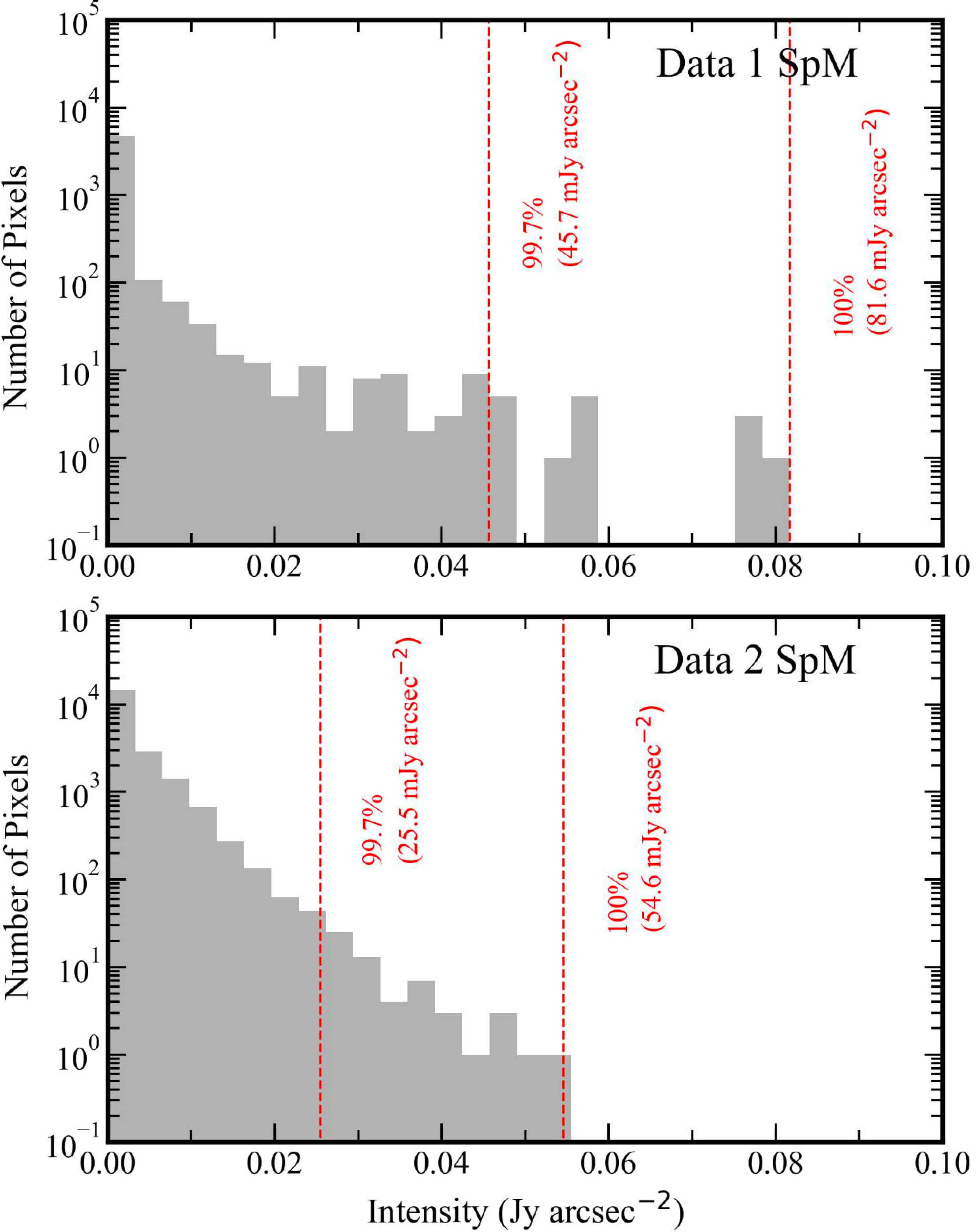}
\caption{Histograms of artificial emissions outside the outer disk (i.e., off-source area) on the raw (i.e., non-convolved) SpM images for Data 1 (top panel) and Data 2 (bottom panel). The vertical dashed lines denote the $99.7\%$ intensity ($\rm I_{99.7}$) and the maximum ($100\%$) intensity ($\rm I_{100}$) of the off-source area.}
\label{fig:offsrc}
\end{figure}

We evaluate the noise level of the raw reconstructed images for Data 1 and 2, respectively. Figure \ref{fig:offsrc} shows a histogram that details the artificial emission outside the outer disk, $2.0$ asec from the phase center (hereafter, the off-source area). Due to both systematic errors and thermal noise, both raw SpM images suffer from artificial emissions in the off-source area. Because of the non-negative constraints in the imaging algorithm, as described in Section \ref{sec:2.3}, the histogram is on the positive side and has a longer tail than that of a Gaussian distribution. A maximum intensity ($\rm I_{100}$) of $81.6$ and $54.5$~mJy~asec$^ {- 2} $ is apparent for Data 1 and 2, respectively. 

Another way to estimate the noise levels is to measure the standard deviation of the residual map, which can be obtained by two-dimensional Fourier transform of the residual visibilities between the model data (which can be obtained by inverse two-dimensional Fourier transform of the reconstructed image) and the observed data.

The residual maps of SpM were reconstructed from the residual visibilities using the DIFMAP software \citep{Shepherd1994}. To minimize the effects caused by frequency-dependent residual gains in the visibility amplitudes, we first made the residual visibilities from the SpM image reconstructed at each spectral window. Then, the residual maps were created through two-dimensional Fourier transform adopting a natural weighting, providing synthesized beams of $0.57\times0.49$ asec with a P.A. = $63^{\circ}$ and $0.22\times0.16$ asec with a P.A. = $80^{\circ}$ for Data 1 and 2, respectively. Finally, we combined the residual maps of all the spectral windows to obtain a final image. The residual maps of MS-CLEAN were generated on the process of the image reconstruction as described in the Section \ref{sec:2.2}. The degree of difference of the beam size on the residual map between the SpM and MS-CLEAN was derived to be $12~\%$.

\begin{figure*}[h]
\centering
\includegraphics[width=0.98\textwidth]{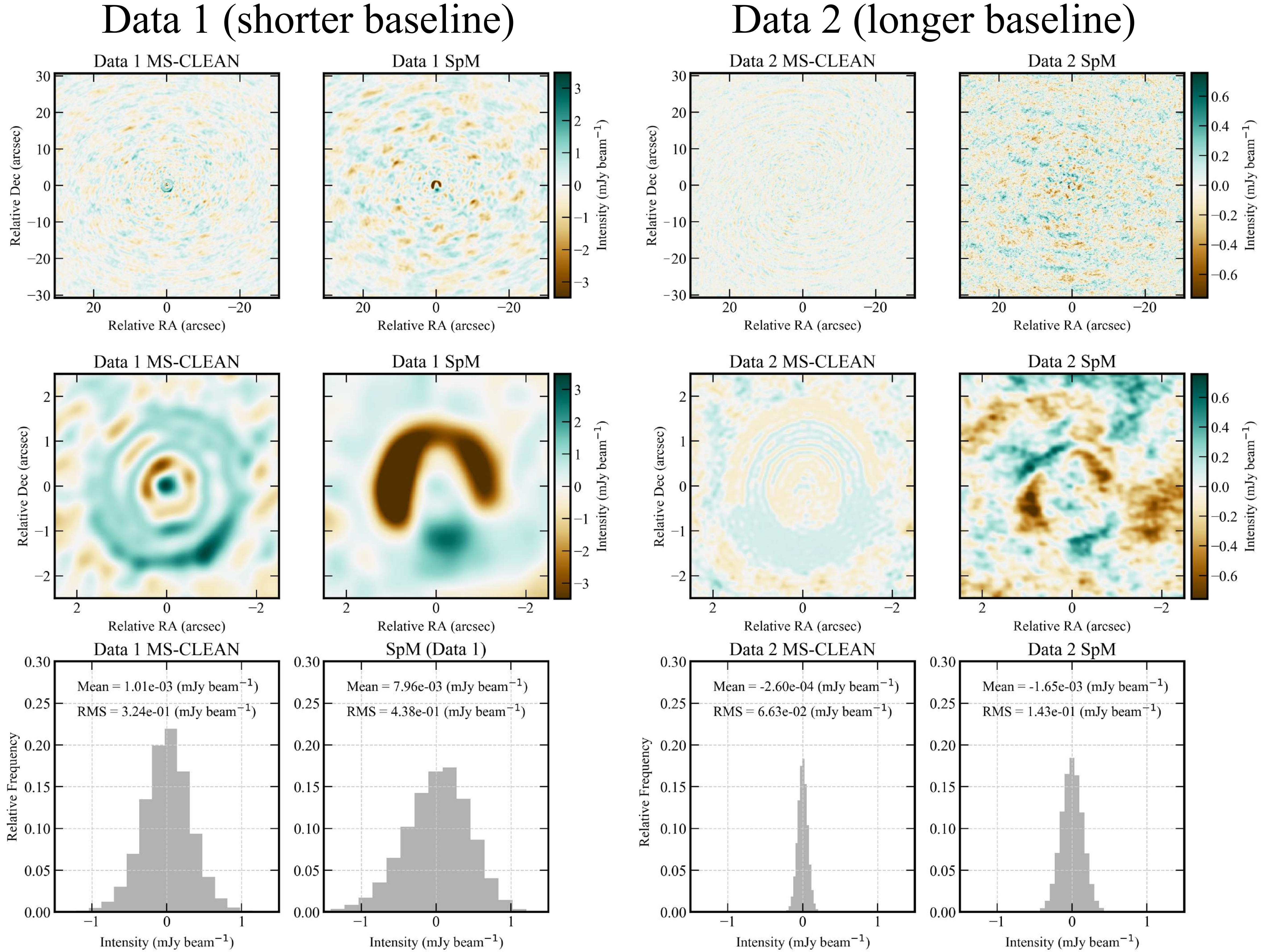}
\caption{Residual maps of the four reconstructed images produced by SpM and MS-CLEAN. The field of view of 61.44 $\times$ 61.44 asec and 5.0 $\times$ 5.0 asec are adopted for the four images in the top and middle panels, respectively. The lower panels show that the off-source regions, which are outside a radius of 2.0 asec from the center coordinates, are extracted from the noise histograms in these images and count pixel values comparable to noise levels in the image domain. The residual maps with MS-CLEAN for Data 1 and Data 2 are convolved with a synthesized beam of $0.51\times0.44$ asec (P.A. = $59^{\circ}$) and $0.20\times0.14$ asec (P.A. = $78^{\circ}$), respectively (see Section \ref{sec:2.2} for details). The residual maps with SpM for Data 1 and Data 2 are convolved with a synthesized beam of $0.57\times0.49$ asec (P.A. = $63^{\circ}$) and $0.22\times0.16$ asec (P.A. = $80^{\circ}$), respectively (see Section \ref{sec:4_noise} for details).}
\label{fig:residual_map}
\end{figure*}

The top and middle panels of Figure \ref{fig:residual_map} show the residual maps of MS-CLEAN and SpM for Data 1 and 2. In the on-source area, MS-CLEAN images have a near symmetrical distribution of the residuals, while SpM images have asymmetric and more residuals. This can also be seen in the off-source area; the SpM residual images have RMS noise of $ 0.44$ and $0.14$ mJy beam$^{-1}$ for Data 1 and 2, respectively, while the MS-CLEAN residual maps show smaller RMS noise of $0.32$ and $0.07$ mJy beam$^ {- 1}$. This is primary because SpM imaging is performed on self-calibrated data for the MS-CLEAN images, and the gains are not precisely solved for SpM images. This is because the SpM imaging solves the observational data by assuming that a modeled object is composed of various smooth scale sizes, while the self-calibration solves gains by assuming that a model is a collection of point sources, adopting a multi-scale approach with the MS-CLEAN algorithm. The processing may lead to both artificial emissions and higher residuals on the SpM image than on the MS-CLEAN image of almost the same resolution.

We regard higher intensity than detection threshold ($>\rm I_{100}$) as the dust emissions from the disk. When we take the beam-convolution into account, $\rm I_{100}$ are comparable to 7 times higher than the RMS noise level in the residual map of Data 1 and 2.7 times for Data 2. Therefore, the emissions above the $\rm I_{100}$ level is likely able to capture the previously known disk structure around HD~142527.


\begin{figure*}[t]
\centering
\includegraphics[width=1.0\textwidth]{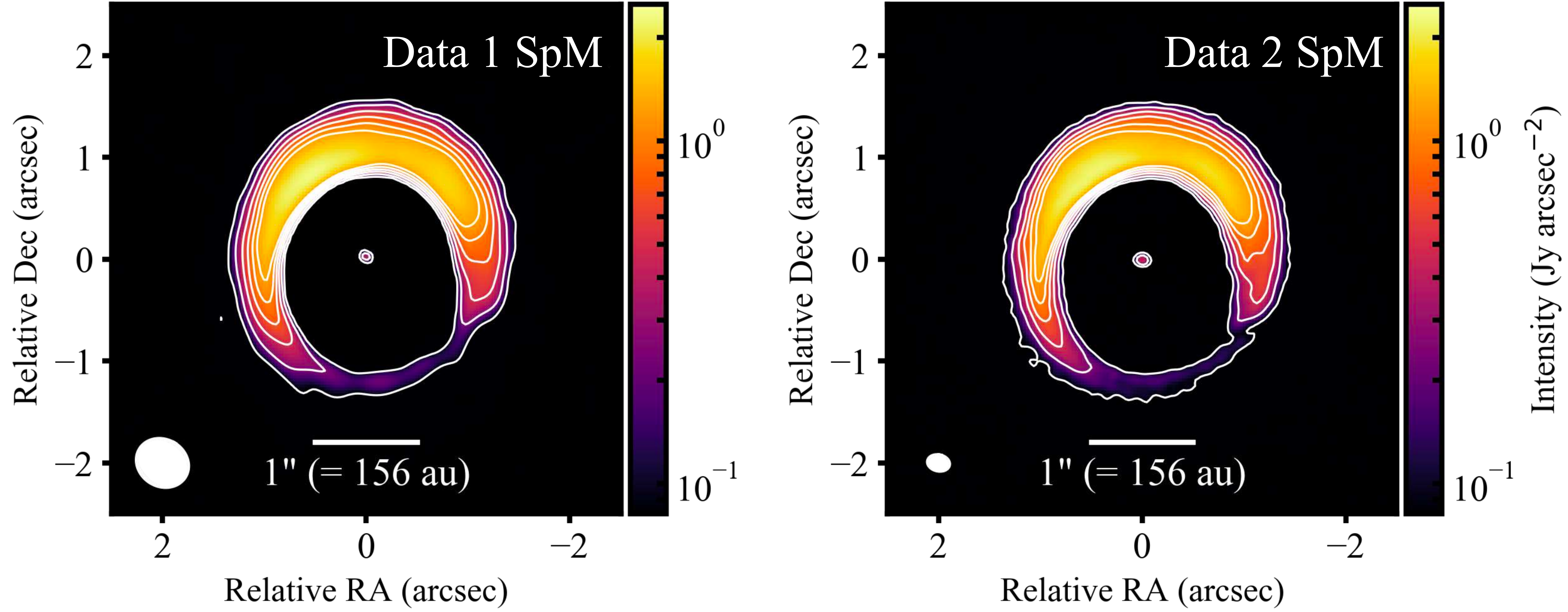}
\caption{SpM images of protoplanetary disk HD~142527 from Data 1 ($left~panel$) and Data 2 ($right~panel$). The same logarithmic color scale and field of view of 5.0 $\times$ 5.0 asec are adopted for images. A white bar of 1~asec ($= 156$~au) is provided for reference to the angular scales. Each contour corresponds to $(1, 3, 6, 9, 12, 15) \times \rm I_{100}$ for Data 1 and $\rm I_{100}$ for Data 1 is 81.6~mJy asec$^{-2}$, as derived in Figure \ref{fig:offsrc}. $\rm I_{100}$ for Data 2 is 54.6~mJy asec$^{-2}$. The total flux of the Data 2 SpM image is scaled to that of the Data 1 SpM image to minimize the effects from flux calibration errors, and these images are not convolved with the Gaussian beam. The beam sizes of nominal resolution of Data 1 and 2 are plotted on its panels respectively to compare the degree of the spatial resolution between the nominal resolution domain and the raw image domain.}
\label{fig:spmim}
\end{figure*}

\subsubsection{Double Ridge-line shown in the Outer Disk}\label{sec:4_dr}

\begin{figure*}
\centering
\includegraphics[width=1\textwidth]{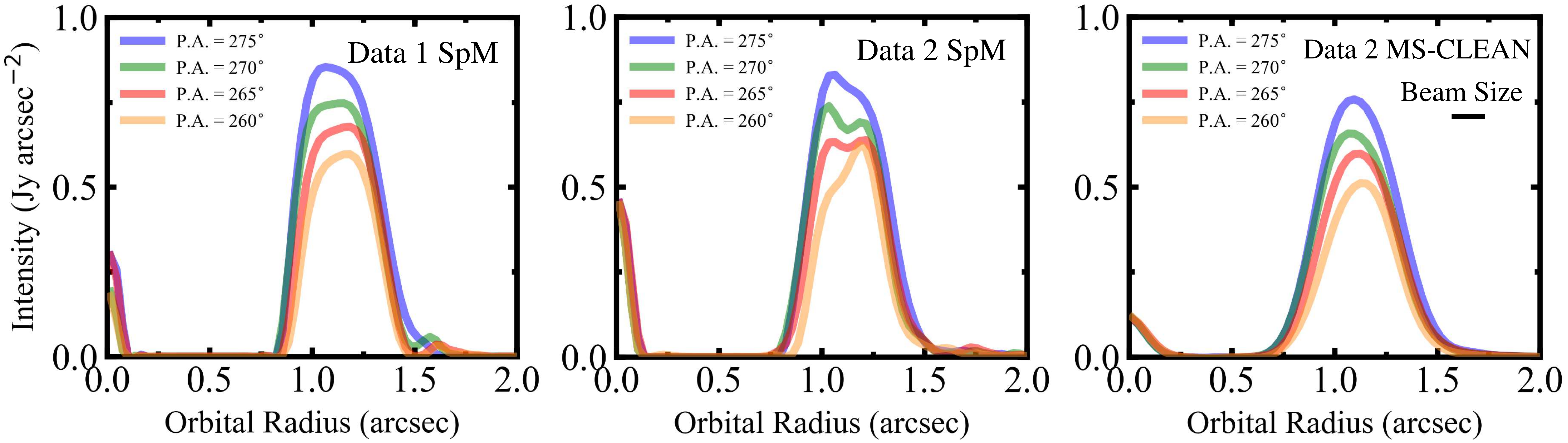}
\caption{The radial profiles along the specified position angles near the break area (P.A. = $260^{\circ}, 265^{\circ}, 270^{\circ}, 275^{\circ}$) for the raw (i.e. non-convolved) reconstructed SpM images for Data 1 and Data 2 (left and middle panels) and the Data 2 MS-CLEAN image (right panel). The horizontal axis indicates the orbital radius starting from the phase center (corresponding to the position of the central star). The vertical axis indicates the intensities of dust emissions. The total flux of the SpM image and MS-CLEAN image for Data 2 are scaled to that of the Data 1 SpM image to minimize the effects from flux calibration errors.}
\label{fig:radial_profile}
\end{figure*}

Figure \ref{fig:spmim} shows the raw reconstructed SpM images of Data 1 and 2 with contours starting at $\rm I_{100}$. Only in the Data 2 SpM image, we found there is a break area of the emission, where the $\rm I_{100}$ contour twists and connects to the bright region toward the north. Moreover, the intensity distribution at P.A. of  $265^{\circ} - 270^{\circ}$ shows a double ridge-like structure. Figure \ref{fig:radial_profile} shows the radial profiles along the specific position angles of the raw reconstructed SpM images as well as those of the Data 2 MS-CLEAN image. The radial profile of the raw Data 2 SpM image only shows a double ridge-like structure at P.A.= $265^{\circ} - 270^{\circ}$, which might indicate the presence of substructure in the horseshoe dust distribution.

We caution that there is so far no clear evidence to present the robust degree of the super-resolution of Data 2 for SpM. The degree of super-resolution that we can achieve for Data 2 varies depending on the signal-to-noise ratio, as well as the $uv$-coverage. Figure~\ref{fig:uv-cov} actually shows that the higher spatial frequency components of $uv$-coverage for Data 2 are relatively sparse. Due to the lower density of the high spatial frequency components, the shorter-baseline data is more weighted in the minimization of Equation \ref{spm_eq}. Such a data set could at least prevent the factor 3 improvement of the spatial resolution with SpM imaging. To assess the consistency of the SpM imaging and definitely confirm the existence of this disk substructure, the observations at higher spatial resolution and deeper sensitivity comparable to DSHARP \citep{Andrews2018}  should be conducted. We also note that more source-specific modeling of brightness distribution on the sky and observational simulations may be necessary to verify the presence of very small scale structures, in a similar manner as \citet{eht2019d}.


\subsection{Future Prospects for Imaging}\label{sec:4_future}
In Section \ref{sec:3}, we have demonstrated that the application of SpM provides high-fidelity super-resolution images. The major structures are in good agreement with a factor of $\sim 3$ times the nominal resolution used in MS-CLEAN imaging. Although this factor is broadly consistent with previous research \citep{honma2014,akiyama2017a, akiyama2017b,Kuramochi2018}, it could be profoundly different for other data sets with different intensity distributions, $uv$-coverages, and/or sensitivities. In future work, we will investigate how to determine the effective resolution for this technique.

We next describe several issues which can be further explored in future work, to improve the signal-noise ratio (SNR) further and the fidelity of reconstructed images. Primary features that limit the SNR of both SpM images are the compact noises seen on the off-source area. Because the locations of these compact noises are broadly consistent with those within the MS-CLEAN components, this could predominantly be due to the miscalibration of the complex gains caused by the self-calibration with MS-CLEAN images. A straightforward way to reduce these compact noises is iterative SpM imaging involving self-calibration. Recent work in \citet{chael2018} has suggested that self-calibration with images from new imaging techniques may significantly reduce such artifacts. \citet{eht2019d} has also adopted this strategy using the SpM imaging produced during this study.

Another factor, which is less important but which can limit the fidelity of our SpM images, are errors coming from the $uv$-gridding. Because the SpM images are produced from $uv$-gridded visibilities, our imaging process is equivalent to a single minor cycle in the Cotton-Schwab CLEAN \citep{schwab1984}. This issue can be mitigated by switching the Fourier transform algorithm to non-uniform FFT (NuFFT) algorithms, which may adopt an image from $uv$-gridded data as the initial model to minimize the number of iterations and NuFFT operations. The major cycle could also be included to reduce the number of NuFFT operations further. This can be achieved by (1) computing residual visibilities on the original $uv$-coordinates with NuFFT using the previous minor-cycle image, (2) deriving $uv$-gridded visibilities by adding $uv$-gridded residual visibilities to the model visibilities from the minor-cycle image, (3) imaging with $uv$-gridded visibilities, and (4) repeating (1-3). 

Finally, we note that for the community to use such new imaging techniques, the implementation of these algorithms in a major software package is essential. The SpM imaging algorithms presented in this paper are currently being implemented as an external module of {\tt CASA} \citep[{\tt PRIISM};][]{nakazato2019}, with many improvements to the core imaging code such as the use of FFT/NuFFT algorithms and with further acceleration of the numerical codes.

\section{Conclusion}\label{sec:5}
We present images of the protoplanetary disk around HD~142527 obtained with ALMA using SpM, which is a new high-fidelity super-resolution imaging technique for radio interferometry. We summarize our main conclusions as follows.

(1) SpM drastically improves the image fidelity of observations of the disk structure around HD~142527 in the super-resolution regime. The Data 1 SpM image achieves an optimal beam size of ~$35-40~\%$ of the nominal resolution for Data 1 by using NRMSE analysis regarding the Data 2 MS-CLEAN image as the reference image. This result means that the Data 1 SpM image achieves $\sim3$ times the higher angular resolution with respect to the nominal resolution of Data 1.

(2) To evaluate the raw (i.e., not convolved) SpM image, we conservatively introduce the detection threshold $\rm I_{100}$, which is the maximum intensity in the emission-free area. With this threshold, we found that new substructures in the horseshoe dust disk are inferred. There is a break area at P.A. of $\sim 230^{\circ}$, where the $\rm I_{100}$ contour twists and connects to the bright region toward the north. Moreover, the intensity distribution at P.A. of $265^{\circ} - 270^{\circ}$ shows a double ridge-like structure. To confirm these notable substructures, the HD 142527 disk deserves a follow-up observation at higher spatial resolution and deeper sensitivity.

(3) SpM images have more asymmetric and larger residuals, while MS-CLEAN images have nearly symmetric residuals distributions. This is predominantly caused by the miscalibration of the complex gains due to the self-calibration with MS-CLEAN. A straightforward of mitigating this problem is iterative SpM imaging involving self-calibration. By combining the implementation of the NuFFT algorithm, SpM imaging will lead to further improvements in the fidelity and SNR of the reconstructed images of protoplanetary disks observed with ALMA. We plan to investigate this in the near future further.

Our results demonstrate that on-going intensive developments of new imaging techniques using SpM is an attractive choice to provide a high-fidelity super-resolution image with ALMA.

\acknowledgements
We thank the anonymous referee, who gave us invaluable comments to improve the paper. We thank Editage (www.editage.com) for English language editing. We thank H. Nagai for the instruction concerning ALMA data reduction, all of the East Asian ALMA staff members at NAOJ for their kind support. MY thanks Sai Jinshi and Toshiki Saito for the helpful conversation. MY is financially supported by a Public Trust Iwai Hisao Memorial Tokyo Scholarship Fund and the ALMA Japan Research Grant of NAOJ ALMA Project, NAOJ-ALMA-229. KA is a Jansky Fellow of the National Radio Astronomy Observatory. The National Radio Astronomy Observatory is a facility of the National Science Foundation operated under cooperative agreement with the Associated Universities, Inc. The Black Hole Initiative at Harvard University is financially supported by a grant from the John Templeton Foundation. This work was financially supported in part by a grant from the National Science Foundation (AST-1614868; KA), JSPS KAKENHI Grant Numbers 
17H01103, 18H05441, and 19K03932 (MT), and 17K14244 (TT), and the Bilateral Joint Research Projects of JSPS (KT). This paper makes use of the following ALMA data: \\
ADS/JAO.ALMA$\#2012.1.00631$.S, \\
ADS/JAO.ALMA$\#2015.100425$.S. \\
ALMA is a partnership of ESO (representing its member states), NSF (USA) and NINS (Japan), together with NRC (Canada), MOST and ASIAA (Taiwan), and KASI (Republic of Korea), in cooperation to the Republic of Chile. The Joint ALMA Observatory is operated by ESO, AUI/NRAO, and NAOJ." The data analysis was in part carried out on the Multi-wavelength Data Analysis System operated by the Astronomy Data Center (ADC), National Astronomical Observatory of Japan.

\facility{ALMA}
\software{Astropy \citep{astropy2013,astropy2018}, CASA \citep{mcmullin2007}, DIFMAP \citep{Shepherd1994}, matplotlib \citep{Hunter2007}}

\restartappendixnumbering  

\appendix
\section{Evaluation of Radial and Azimuthal Structure of the Outer Disk: In the Case of Data 2 SpM Image}\label{appendix_a}
\begin{figure}[t]
\centering
\includegraphics[width=0.48 \textwidth]{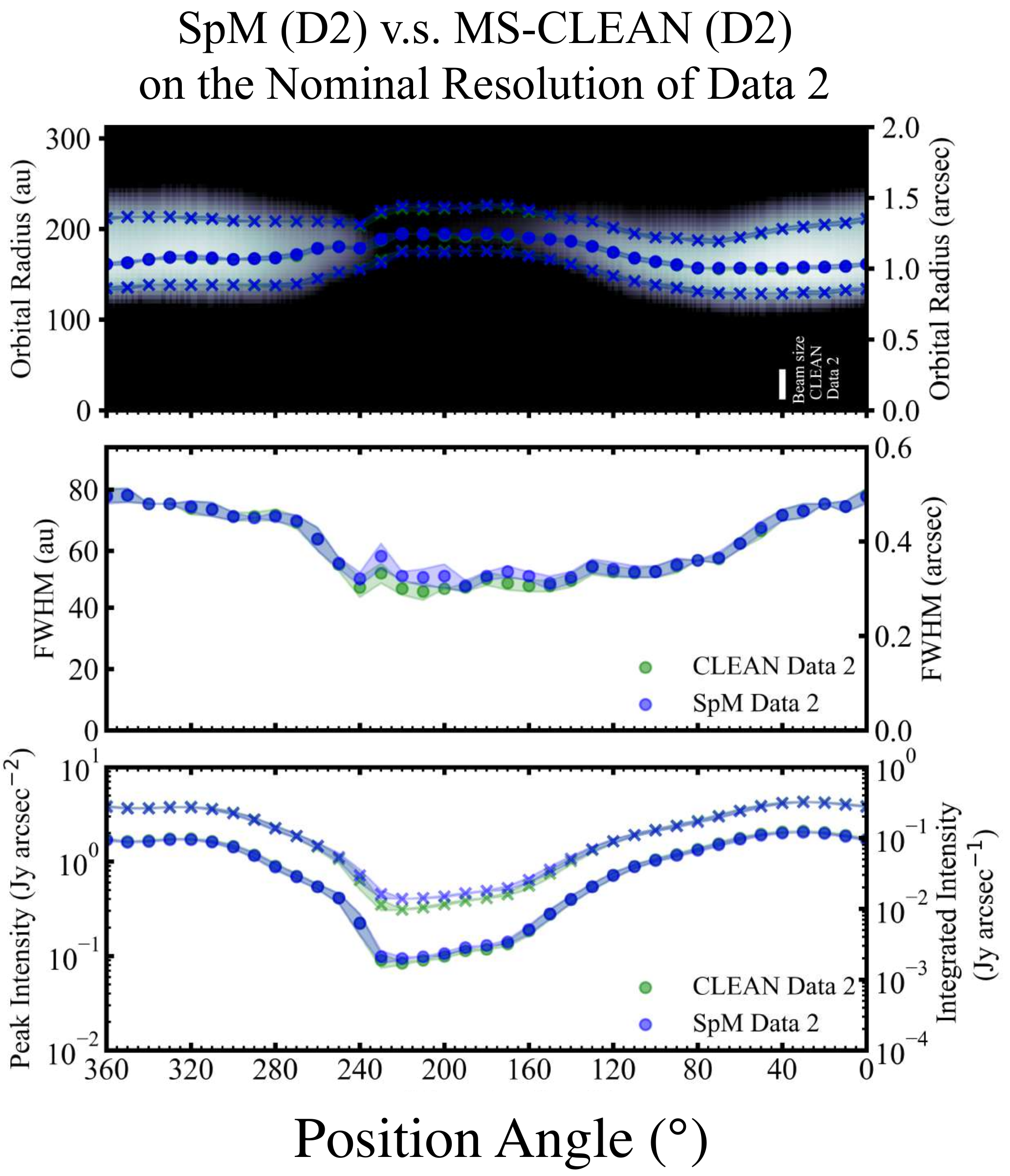}
\caption{The position angle (P.A.) profile for the Data 2 SpM (blue color) and Data 2 MS-CLEAN (green color) images convolved with the Data 2 MS-CLEAN nominal resolution. The top panel shows the radial location of the peak (circle points) and its $50\%$ (i.e., FWHM; cross points) of the intensity distribution of the outer disk overlaid by the Data 2 MS-CLEAN image. The middle panel shows the radial FWHM sizes of the intensity distribution for the outer disk. Bottom panel shows the peak (circle points) and radially integrated (cross points) intensities.}
\label{fig:pa_profile_2}
\end{figure}

Figure \ref{fig:pa_profile_2} shows the peak and the FWHM of the radial surface brightness profiles at position angles for the Data 2 SpM and Data 2 MS-CLEAN images convolved with the nominal resolution of Data 2. The profiles are in the same manner as Figure \ref{fig:pa_profile}. The total flux of the Data 1 SpM image is scaled to that of the Data 2 MS-CLEAN image to minimize the effects from errors in the flux calibration.

Table \ref{tab:pa_residuals_2} summarizes the residual statistics for the physical parameters of the P.A. profiles, as shown in Figure \ref{fig:pa_profile_2}. The residual parameters are subtracted from the two images on each P.A. profile. The radial locations of the peak and its halves (outer/inner half peak) and the radial FWHMs are mostly consistent within nearly 10$\%$ of the nominal angular resolution of Data 2, which is close to the pixel size of the image. The peak and integrated intensity are also consistent; the mean and standard deviation are comparable to the noise levels on the residual maps estimated in Section \ref{sec:4_noise}. Therefore, Figure \ref{fig:pa_profile_2} indicates that each profile is in good agreement in terms of the radial and azimuthal structure of the outer disk.

\begin{table}[ht]
\caption{Residual Statistics of P.A. profiles}
\label{tab:pa_residuals_2}
\begin{tabularx}{\linewidth}{lccccccccc}
\toprule
& \multicolumn{3}{c}{SpM (D2) - MS-CLEAN (D2)}\\
\cmidrule(lr){2-4} \cmidrule(lr){5-7} \cmidrule(lr){8-10}
Quantities (unit)              & Mean     & Std     & Abs.Max   \\ \hline
Radial Peak (mas)              & 3.00     & 4.30    & 12.00    \\
Outer Half Peak (mas)          & -1.95     & 4.26    & 3.00    \\
Inner Half Peak (mas)          & 4.38     & 6.69    & 21.00    \\
FWHM (mas)                     & 6.32     & 10.32    & 36.00     \\
Peak I $\mathrm{(mJy~asec}^{-2})$  & -19.43    & 24.35    & 11.05  \\
Integr. I $\mathrm{(mJy~asec}^{-1})$ & -1.22     & 4.46    & 4.68 \\
\bottomrule
\end{tabularx}
\end{table}

\section{Fidelity at Multi-resolution using NRMSE regarding Data 2 SpM Image as the reference image}\label{appendix_b}

We present the results of NRMSE analysis by regarding the raw (i.e., non-convolved) Data 2 SpM image as the reference image. Table \ref{tab:beamsize_d2spm} summarizes the optimal resolutions as determined by the NRMSE. For each image, the worst optimal resolution was selected from the results of two NRMSE analyses (for the image and gradient domains) and defined as the optimal resolution. In the following paragraphs, we describe in more detail the characteristics of each curve (image and gradient domain) and their relation to the corresponding images.

The Data 1 MS-CLEAN image shows significant deviations from the reference image at R = $50-60~\%$. This achieves the optimal resolution at $55~\%$ and sharply increases the NRMSEs at finer resolutions. The compact artifacts created in MS-CLEAN dominate the deviations from the other lines for NRMSE.  The Data 2 MS-CLEAN starts to deviate from the SpM image for the same data set immediately after the resolution becomes smaller than nominal. The optimal resolution achieves at R = $25~\%$, corresponding to $70~\%$ of the Data 2 nominal resolution. Similar to Data 1, the Data 2 MS-CLEAN image shows a rapid increase in the NRMSE at finer resolutions, which is dominated by small artifacts. Therefore, the NRMSEs of the MS-CLEAN images have similar behavior due to compact artifacts that appear clearly at $\rm R \leq 50-60~\%$. The above results are consistent with previous work on imaging simulations \citep{chael2016,akiyama2017a,akiyama2017b, Kuramochi2018}.

In contrast, the Data 1 SpM image follows the curves of the reference image until at R  = $30-40~\%$ in both the image and gradient domains. The optimal resolution reaches R = $15~\%$, which is better than the images from MS-CLEAN. At resolutions of $\rm R \leq 20~\%$, the NRMSE of the Data 1 SpM shows flat curves, with typical ranges of $5-10~\%$ on the image domain and $20-25~\%$ on the gradient domain. It can be assumed that the NRMSE of $5-10~\%$ on the image domain is caused by systematic errors between two data sets. In particular, the self-calibrated visibilities that we used could be the origin of the systematic errors, because they were iteratively calibrated using the MS-CLEAN images.

Figure \ref{fig:residual_spmimage} shows the Data 1 SpM images at the optimal resolution and without any post-processing Gaussian, and a residual image between both images, respectively. As shown in the residual image, residual intensities are less than $10-20\%$ of the original images, indicating that the images are almost in agreement. This means that the intensity distribution could be resolved even with the optimal resolution, leading the flatness in the NRMSE curve.

\begin{figure*}[h]
\centering
\includegraphics[width=1.0\textwidth]{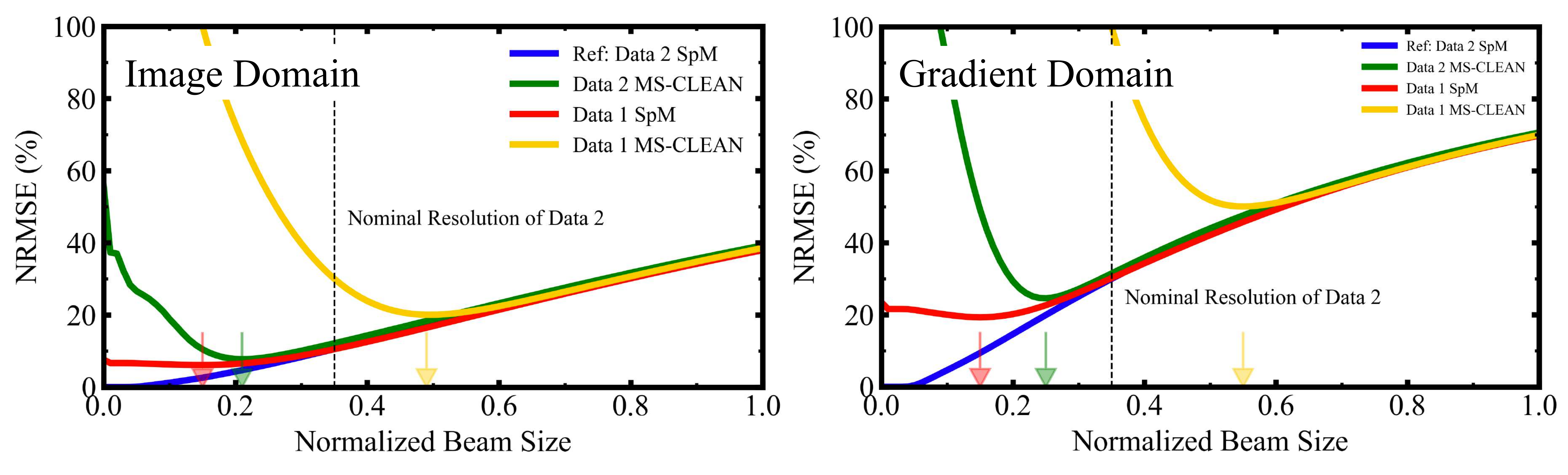}
\caption{
NRMSEs of the four reconstructed images as a function of the normalized beam size, on the image domain (the left panel) and the gradient domain (the right panel). The raw (i.e., non-convolved) Data 2 SpM image is adopted as the reference image for the NRMSE analysis in both panels. Each image is convolved with an elliptical beam size, with the same axial ratio and P.A. as the nominal resolution of Data 2. The nominal resolution of Data 1 normalizes the size of the convolving beam shown in the horizontal axis.
The solid blue line represents the NRMSE between the non-convolved and convolved SpM images from Data 2 , producing the ideal case in which there is no difference between the input and reference images. The red, yellow, and green lines represent the NRMSE curves for the SpM image from Data 1 , and the MS-CLEAN images for Data 1 and 2, respectively.
The red, green, and yellow arrows indicate the resolutions minimizing the NRMSE curves with the same colors, which can be interpreted as the optimal beam sizes of the corresponding images. The vertical dashed line indicates the nominal resolution of Data 2, which is $35~\%$ (= $20\times14$ mas) of Data 1. 
}
\label{fig:nrmse_d2spm}
\end{figure*}

\begin{figure*}[ht]
\centering
\includegraphics[width=1\textwidth]{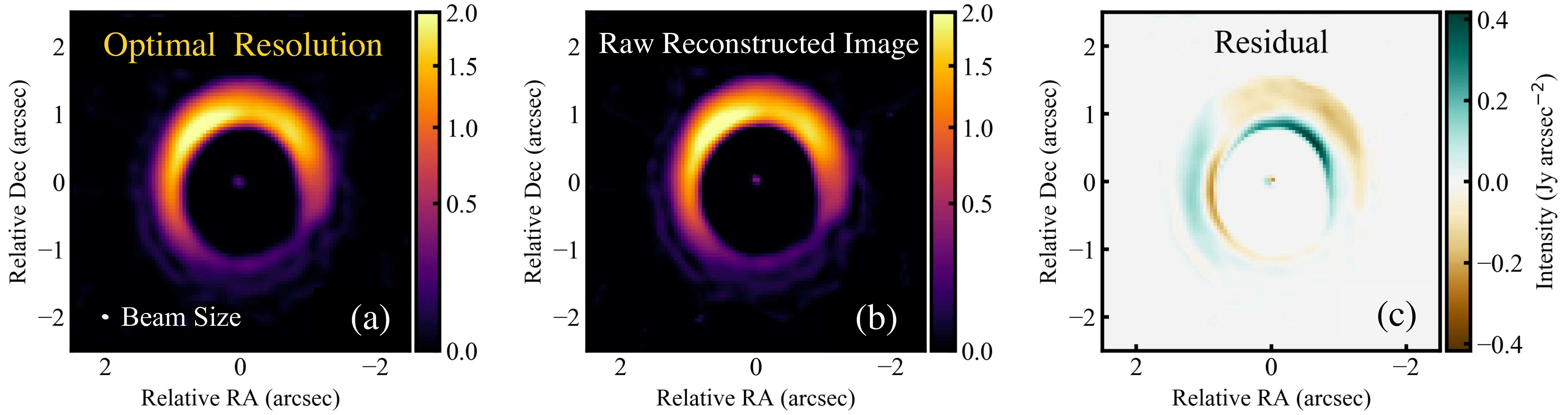}
\caption{Panel (a) shows the reconstructed image convolved with the optimal resolution of Data 1 for SpM as determined by NRMSE analysis by regarding the raw SpM image from Data 2 as the reference image (see Figure \ref{fig:nrmse_d2spm}). The optimal resolution is $15~\%$ (= $9\times6$ mas) of the nominal resolution for Data 1, which is shown in a white ellipse on the bottom left corner. Panel (b) shows the raw (i.e., non-convolved) Data 1 SpM image. Panel (c) shows the residual image $\left\{(a)-(b)\right\}$ between the panel (a) and (b) on the intensity distribution.}
\label{fig:residual_spmimage}
\end{figure*}

The results shown in Figure \ref{fig:nrmse_d2spm} suggest that the traditional post-processing Gaussian convolution does not help to improve the image fidelity of SpM images, while it is definitely required to smooth the compact artifacts attributed to the underlying assumption and to improve the fidelity of images produced via MS-CLEAN. With the adoption of multi-resolution regularization using TSV, a piecewise smooth image that maintains the consistency with the observed visibilities can be reconstructed using SpM. This enables high-fidelity imaging, even without convolution with a Gaussian beam. The significant advantage of using such multi-resolution regularization has already been demonstrated in previous investigations involving such simulation \citep{akiyama2017a,akiyama2017b, Kuramochi2018,birdi2018}.

It should finally be noted that the unnecessity of using post-processing convolution in SpM imaging does {\it not} mean that the reconstructed image has an infinite angular resolution. There should be a certain limit to the angular resolution where data sets do not significantly constrain the structure and regularization functions do not effectively work to constrain it. The previous study has reported that SpM could separate a two-point source at a resolution corresponding to a beam size of $25-40~\%$ of the beam size for the nominal resolution \citep[e.g.,][]{honma2014, Kuramochi2018}. In the current study, the effective resolution for this technique is $30-40\%$ of the beam size for the nominal resolution, where the NRMSE curves start to deviate from those of the reference image. Although this factor is broadly consistent with previous research, this could be profoundly different for other data sets with different uncertainties and $uv-$coverages or source structures. The evaluation of the level of resolution improvement for such data sets is in the scope of our next studies in the near future.

\begin{table*}[h]
\caption{Optimal resolutions determined by NRMSE: in the case of Data 2 SpM image as the reference image} 
\label{tab:beamsize_d2spm}
\begin{tabularx}{\linewidth}{lrrr}
\toprule
Angular Resolution & SpM (Data 1) &MS-CLEAN (Data 1) & MS-CLEAN (Data 2) \\
\hline
Nominal Resolution (mas) & NaN & 57 $\times$  40~(100~$\%$) & 20 $\times$ 14~(35~$\%$) \\
Optimal Resolution on Image Domain (mas) & 9 $\times$ 6~(15~$\%$) & 28 $\times$ 20~(49~$\%$) & 12 $\times$ 8~(21~$\%$) \\
Optimal Resolution on Gradient Domain (mas) & 9 $\times$ 6~(15~$\%$) & 31 $\times$ 22~(55~$\%$) & 14 $\times$ 10~(25~$\%$) \\
\bottomrule
\end{tabularx}
\tablecomments{The axial ratio and the P.A. of the beams for Data 1 are fixed to those of the synthesized beam size of the MS-CLEAN image for Data 2. The percentages indicate the beam size ratio for the nominal resolution of Data 1.}
\end{table*}



\bibliographystyle{aasjournal}
\bibliography{ref}
\end{document}